\documentclass[fleqn,usenatbib]{mnras}

\usepackage[T1]{fontenc}

\DeclareRobustCommand{\VAN}[3]{#2}
\let\VANthebibliography\thebibliography
\def\thebibliography{\DeclareRobustCommand{\VAN}[3]{##3}\VANthebibliography}

\usepackage{graphicx}	
\usepackage{amsmath}	
\usepackage{amssymb}	
\usepackage{bigdelim}   
\usepackage{booktabs}   

\usepackage{xcolor}     
\usepackage{newtxtext,newtxmath}

\def\msun{{\rm ~M}_{\odot}}
\newcommand{\kms}{{\rm {km\, s^{-1}}}}

\newcommand{\comment}[1]{} 

\newcommand\mc[1]{\multicolumn{1}{c}{#1}}

\title[IMBHs escaping from GCs]{MOCCA-SURVEY Database II -- Properties of Intermediate Mass Black Holes escaping from star clusters}

\author[K. Maliszewski et al.]{
Konrad Maliszewski,$^{1}$
Mirek Giersz,$^{2}$\thanks{E-mail: mig@camk.edu.pl}
Dorota Gondek-Rosińska,$^{1}$
Abbas Askar,$^{3}$
Arkadiusz Hypki$^{2}$
\\
$^{1}$Astronomical Observatory Warsaw University, 00-478 Warsaw, Poland\\
$^{2}$Nicolaus Copernicus Astronomical Centre, Polish Academy of Sciences, Warsaw, Poland\\
$^{3}$Lund Observatory, Department of Astronomy, and Theoretical Physics, Lund University, Box 43, SE-221 00 Lund, Sweden
}

\date{Accepted XXX. Received YYY; in original form ZZZ}

\pubyear{2021}

\begin{document}
\label{firstpage}
\pagerange{\pageref{firstpage}--\pageref{lastpage}}
\maketitle

\begin{abstract}
In this work we investigate properties of intermediate-mass black holes (IMBHs) that escape from star clusters due to dynamical interactions. The studied models were simulated as part of the preliminary second survey carried out using the MOCCA code (MOCCA-SURVEY Database II), which is based on the Monte Carlo \textit{N}-body method and does not include gravitational wave recoil kick prescriptions of the binary black hole merger product. We have found that IMBHs are more likely to be formed and ejected in models where both initial central density and central escape velocities have high values. Most of our studied objects escape in a binary with another black hole (BH) as their companion and have masses between $100$ and $\rm 140 \msun$. 
Escaping IMBHs tend to build-up mass most effectively through repeated mergers in a binary with BHs due to gravitational wave emission. Binaries play a key role in their ejection from the system as they allow these massive objects to gather energy needed for escape. The binaries in which IMBHs escape tend to have very high binding energy at the time of escape and the last interaction is strong but does not involve a massive intruder. 
These IMBHs gain energy needed to escape the cluster gradually in successive dynamical interactions. We present specific examples of the history of IMBH formation and escape from star cluster models. We also discuss the observational implications of our findings as well as the potential influence of the gravitational wave recoil kicks on the process.
\end{abstract}

\begin{keywords}
stars: black holes -- methods: numerical -- globular clusters: general
\end{keywords}



\section{Introduction}


Black holes (BHs) with masses in the range $\rm 10^2 - 10^5 \msun$ are known as intermediate-mass black holes (IMBHs). This range defines BHs that are more massive than stellar-mass BHs and less massive than supermassive black holes (SMBHs). 
The LIGO/Virgo collaboration recently released the catalogue containing the detection of 8 binary black hole (BBH) mergers in which the remnant is an IMBH \citep{gwtc-3}. One such event, also known as GW190521, provided first definite proof for the existence of low-mass IMBHs \citep{2020PhRvL.125j1102A}. Until its discovery, the existence of IMBHs had been heavily debated and concrete evidence for their presence was lacking although several promising candidates \citep[see][and references therein]{greene2019rev} had been observationally identified through different methods \citep[e.g.,][]{2018NatAs...2..656L, 2018ApJ...863....1C, 2018ApJ...859...86T, 2020ApJ...890..167T}.

The two BHs that merged in the GW190521 event had masses of $\sim 85\rm  \msun$ and $\sim 66\rm \msun$. BHs with these masses are not expected to form through the isolated evolution of single massive stars. This is because the progenitor stars that should form BHs with these masses are likely to undergo pair or pulsational pair-instability supernova \citep{Fryer2012,2016A&A...594A..97B}. These supernova events can lead to either enhanced mass loss or the complete explosion of the star leaving behind no remnant. Therefore, they can prevent the formation of BHs with masses in the range $\sim 50 \ \msun$ to $120 \ \msun$, which is referred to as the upper-mass gap \citep{2017ApJ...836..244W,2017MNRAS.470.4739S}. Different formation channels have been proposed to explain the astrophysical origin of the merging BBHs that are being observed by the LIGO/Virgo collaboration, including events like GW190521. Among others, these channels include: 

\begin{enumerate}
    \item dynamical BBH formation in dense stellar environments \citep[e.g.,][]{2000ApJ...528L..17P,2017MNRAS.464L..36A,2018MNRAS.473..909B, 2021arXiv210501085D}. Repeated BH mergers \citep[e.g.,][]{2019PhRvD.100d3027R,2019MNRAS.486.5008A,2020arXiv200609744S,2021arXiv210704639F, 2021arXiv210507003A}, BHs forming from dynamical stellar mergers or binary evolution mergers have been invoked to explain the formation of GW190521 in star clusters \citep[e.g.,][]{dicarlo2020,kremer2020,banerjee2021,2021ApJ...908L..29G}.  
    \item isolated binary star evolution \citet{2013ApJ...779...72D,2016Natur.534..512B, 2021A&A...651A.100O}
    \item mergers in triple and quadruple systems \citep[e.g.,][]{2012ApJ...757...27A, 2020ApJ...903...67M, 2020ApJ...900...16F,2021ApJ...907L..19V}
    \item formation in AGN disks \citep[e.g.,][]{mckernan,2017ApJ...835..165B,2019ApJ...876..122Y,2020ApJ...898...25T,2020arXiv201009765S,2021ApJ...908..194T}
\end{enumerate}

The high stellar densities in environments like globular clusters (GCs) can lead to frequent encounters between stars and binary systems which can result in the dynamical formation of BBHs. Additionally, dynamical interactions in these dense environments may also lead to the formation of IMBHs. GCs have been the subject of extensive theoretical research examining possible IMBH formation mechanisms and the consequences that its presence can have on the evolution of the cluster \citep[e.g.][]{2015MNRAS.454.3150G,2018MNRAS.481.2168M,2019arXiv190500902A,2020MNRAS.498.4287H}. In the cores of initially dense clusters, massive stars may undergo runaway collisions \citep{SPZ02,SPZ04,sakurai2017} that could lead to the formation of a very massive star larger than $\gtrsim 150 \ \msun $ \citep{mapelli2016,reinoso2018}. In low metallicity environments, where mass loss due to stellar winds is low, these massive stars may evolve into IMBH seeds \citep{2017MNRAS.470.4739S,kremer2020}. It is also possible to form IMBH seeds in GCs through the gradual growth of stellar-mass BHs via mergers with other BHs or stars \citep{Miller02,2015MNRAS.454.3150G,2021MNRAS.501.5257R,2021arXiv210501085D}. It has been shown that seed IMBHs of $\rm 10^{2} - 10^{3}\,\rm M_{\odot}$ can grow to larger masses through tidal capture and disruption of stars in extremely dense clusters \citep{alexander2017,stone2017,sakurai2019} with stellar densities $\rm \gtrsim 10^{6} \ \msun \ pc^{-3}$. It has also been suggested that stellar-mass BHs can grow into IMBHs by accreting gas in primordial, gas-rich, massive star clusters \citep{2010ApJ...713L..41V,2013MNRAS.429.2997L}. 

Observationally determining the presence of IMBHs in the crowded centres of GCs is difficult \citep{mezcua2017,devita2017,zocchi2019,aros2020,haberle2021}. Kinematic observations of the central regions of many Galactic GCs have identified upper limits on central IMBH masses \citep{noyola2008,lutzgendorf13,lanzoni13,haberle2021}. However, the results of these works remain inconclusive and there is no conclusive evidence for the presence of an IMBH in the centre of any of the Galactic GCs \citep{mann2019,baumgardt2019}. Observers have also looked for observational signatures of accreting IMBHs at centre of Galactic GCs using X-ray and radio observations \citep{strader2012}. These studies show that is unlikely for Galactic GCs to be harbouring IMBHs more massive than a $\sim 1000 \ \msun$ \citep{tremou18}. The best observational evidence for IMBHs in GCs come from extragalactic observations.  For instance, \citet{lin18} identified an X-ray flare in an extragalactic star cluster that was caused by a tidal disruption event associated with a $\rm 1.75 \times 10^{4} \msun \ pc^{-3}$ IMBH \citep{lin2020,wen2021}. Several IMBH candidates have also been found in central regions of dwarf and low-mass galaxies through dynamical mass measurements \citep{Nguyen2019} and electromagnetic observations \citep{baldassare2015}. Several IMBH candidates have also been identified close to the Galactic center through observations of molecular gas streams \citep{2019PASJ...71S..21T,2020ApJ...890..167T}. It is possible that such IMBHs may have originated in disrupted star clusters that sink towards the Galactic center due to dynamical friction \citep{sedda-gualandris2018,askar2021arxiv}.

If an IMBH forms early during the cluster evolution, it can co-exist in the cluster with several hundreds of stellar-mass BHs \citep{2019arXiv190500902A}. It is highly probable that these stellar-mass BHs will undergo strong dynamical encounters with the IMBH in the cluster center. It is expected that these encounters are likely to eject our stellar-mass BHs from the cluster \citep{Lutz2013,Hong2020MNRAS.498.4287H}. However, it may also be possible to eject out an IMBH in such interactions, particularly when a star cluster hosts multiple BHs with masses larger than $100 \ \msun$. Although the possibility for an IMBH to escape from its host star cluster as a result of dynamical interactions has not been intensively studied, numerical studies indicate this can occur in realistic star clusters \citep{2021arXiv210501085D}.  Several works have investigated the ejection of IMBH and their seeds from star clusters due to gravitational wave recoil kicks following the merger of two BHs \citep[e.g.,][]{2018MNRAS.481.2168M, 2019PhRvD.100d3027R,2021MNRAS.502.2682A, 2021arXiv210507003A, 2021MNRAS.501.5257R}. The magnitude of these gravitational wave recoil kicks depends on the anisotropy in the masses and spins of the merging BHs \citep{gonzalez2007,holley2008}. If these kicks exceed the escape velocity of the cluster than this may result in the ejection of an IMBH seed and hence lower its retention probability \citep{gerosa2019}. Such recoil kicks are are more likely to eject potential IMBH seeds that are less massive than a $100 \ \msun$.

In this paper, we investigate the dynamical ejection of IMBHs ($> 100 \ \msun$) in dense and massive GCs. We present results from star cluster models simulated as part of the MOCCA-SURVEY Database II. In the evolution of some of these models, we find that IMBHs can be ejected from the cluster solely due to dynamical interactions with other objects in the core. The paper is organized as follows. In the section \ref{sec:sec2} we will present the methods used as well as models that were part of the Survey 2. In section \ref{sec:sec3} we will discuss if there is any dependency on the initial parameters that were used in the simulations. The physical parameters of escaping IMBHs will be shown in section \ref{sec:sec4}. In section \ref{sec:sec5} we will present what were the mass build-up events that contributed to the mass of IMBH that was ejected compared to the ones that stay in the system until the Hubble time. The physical processes that are responsible for the IMBHs escapes are discussed in section \ref{sec:sec6}. The most peculiar cases of such escapers are shown in section \ref{sec:sec7}. In section \ref{sec:sec8.5} we try to quantify the effect that the gravitational recoil can have on the ejection process of the IMBHs. Lastly, in section \ref{sec:sec9}, we sum up the most important results of the work while discussing main results and drawing conclusions.



\section{Models} \label{sec:sec2}

The numerical simulations of GCs for the purpose of this work were done with the \textsc{mocca}\footnote{\url{http://moccacode.net}} code \citep{Giersz2014arXiv1411.7603G,Hypki2013MNRAS.429.1221H,Giersz1998MNRAS.298.1239G}. \textsc{mocca} is one of the most feature rich codes which is able to simulate  GCs with realistic sizes. It is much faster than direct \textit{N}-body codes, it follows the cluster evolution very precisely and at the same time it provides almost the same amount of details of the dynamical and stellar evolution of all stars in the cluster \citep{Wang2016MNRAS.458.1450W}. \textsc{mocca} code already proved its value and versatility in a number of projects to simulate the properties of realistic  GCs \citep{Giersz2011MNRAS.410.2698G}, the evolution of compact binaries e.g. BHs binaries \citep{Hong2020MNRAS.498.4287H}. \textsc{mocca} code uses \textsc{sse/bse} codes to advance the stellar evolution of stars and binaries \citep{Hurley2000MNRAS.315..543H, Hurley2002MNRAS.329..897H}. The close dynamical interactions, performed in \textit{N}-body way, are done with \textsc{fewbody} code \citep{Fregeau2004-01-004}. 

Recently, \textsc{mocca} code was used to create a large number of numerical simulations for various initial conditions -- called \textit{mocca} surveys. The newest version of the surveys is MOCCA-SURVEY Database II which an upgrade of the already published and widely used MOCCA-SURVEY Database I \citep{2017MNRAS.464L..36A}. 

The newest feature of the \textsc{mocca} code is the ability to follow the dynamical evolution of the multiple stellar populations \citep{Piotto2015AJ....149...91P}. The initial conditions for MOCCA-SURVEY Database II were chosen to test the code and to test the dynamical mixing between two stellar populations for mostly very dense models, up to central densities larger than $\rm \sim 10^7\: M_{\sun} \ pc^{-3}$. MOCCA-SURVEY Database II currently contains 212 very detailed models which cover a broad range of initial conditions (for details see \citet[in prep.]{Hypki2021}). In this study, the models from the preliminary MOCCA-SURVEY Database II are used. These models will be extended in the near future.  

Among the many initial parameters for \textsc{mocca} models, the one which has importance for the topic of this paper is the flag that determines the remnant mass following neutron star (NS) and BH formation. For MOCCA-SURVEY Database II the rapid and the delayed supernova models from \citet{Fryer2012} were applied in order to have some BH population within the first mass gap range ($2-5$ $\msun$). 
The metallicity for all models is 0.001 (1/20 of the solar metallicity). Models with various metallicities are planned for the next stage of the MOCCA-SURVEY Database II. Wind prescriptions were chosen to be similar to \citet{Belczynski_2010}. The final mass after mergers between NSs/BHs with other stars was modified to be more conservative and now only 25 percent of the merging star's mass is accreted by the compact object. In these models, BH natal kicks are drawn from the same distribution that is used for NSs (\citet{hobbs2005};  Maxwellian distribution with $\rm \sigma = 265 \ km \ s^{-1}$). However, BH kicks are corrected for linear momentum conservation and thus depend on the final BH mass and the fallback prescription defined by the supernova model.  \citet{Fryer2012,Belczynski2016A&A...594A..97B,Banerjee_2020}.
Pulsational pair instability (PPISN) and full-fledged pair-instability supernova (PISN) were chosen to follow \citet{Belczynski2016A&A...594A..97B}. Possibility for the electron capture and accretion induced SNs are also switched on. The effect of the tidal disruptions is not included in the simulated models. The overall comparison of the code with the state-of-the-art \textit{N}-body code \textsc{NBODY6++GPU} and the treatment of the stellar winds can be found in \cite{kamlah}.

\begin{figure}
    \centering
    \includegraphics[width=\columnwidth]{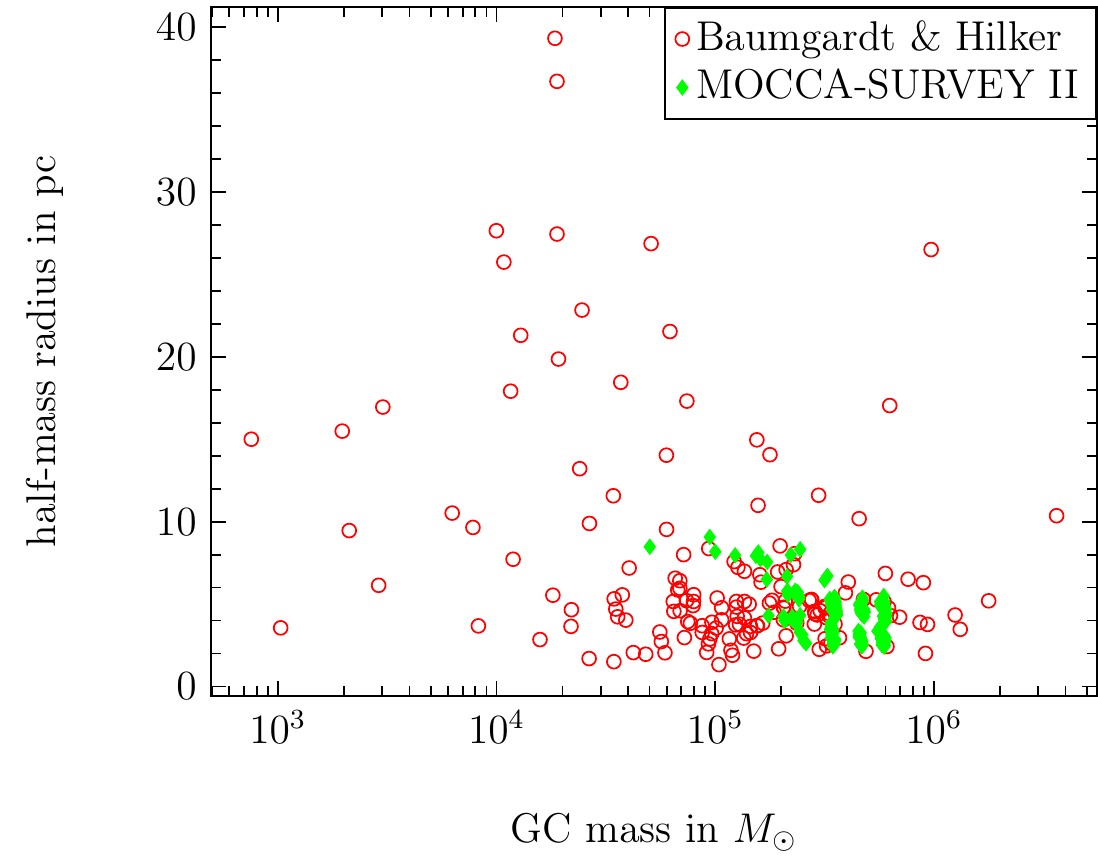}
    \caption{Comparison of the global parameters of GCs produces with the \textsc{mocca} code (green filled diamonds) and the observed ones from the \protect\cite{baumgardt} catalogue (red empty circles). The MOCCA-SURVEY II models are evolved to 12 Gyrs.}
    \label{fig:fig0}
\end{figure}

The best indication on how well does the code work is to compare its output models with the observed GCs. To do that we used the \cite{baumgardt} catalogue to compare the mass and half-mass radius of real GCs to the ones produced by the \textsc{mocca} code. The results are presented in Fig.\,\ref{fig:fig0} and clearly show that our models, which are evolved up to 12 Gyrs, have properties that match the observations. However, it is worth noticing that our models do not cover the whole range of the observed parameters and thus cannot be treated as representative of all the clusters in the Universe. In fact, they represent clusters which are rather dense.

\section{Initial model parameters} \label{sec:sec3}

Firstly, in Table \ref{tab:tab1} we present initial conditions for the models that constituted the MOCCA-SURVEY Database II. In order to determine, on which initial parameters the IMBH formation and its ejection depends, we studied all 212 models available in the MOCCA-SURVEY Database II. We have distinguished models in which: 
\begin{enumerate}
    \item there was at least one IMBH escaper in the model and no IMBH was present in the cluster at 12 Gyr (49 models)
    \item an IMBH was present in the cluster at 12 Gyr and there were no escaping IMBHs during the evolution of the cluster (66 models)
    \item there was at least one IMBH escaper and an IMBH was present in the cluster at 12 Gyr (20 models).
\end{enumerate}
This way we could differentiate between the IMBH formation inside the cluster and processes that lead to their escapes. In Table \ref{tab:tab2}, we present the initial parameters that differ between the simulated star cluster models. For some of the initial parameters, value ranges have been used to group models and Table \ref{tab:tab2} provides the total number of models belonging to each group.

We have found that for our models there was no dependency on the initial number of objects in the cluster for IMBH formation. This is due to the fact that models in the MOCCA-SURVEY Database II took rather extreme initial parameters values, their initial central densities were high and clusters were strongly tidally-underfilling, however, these conditions may correspond to the initial values of the observed clusters \citep[see][]{2022A&A...659A..96M}.

\begin{figure}
    \centering
    \includegraphics[width=\columnwidth]{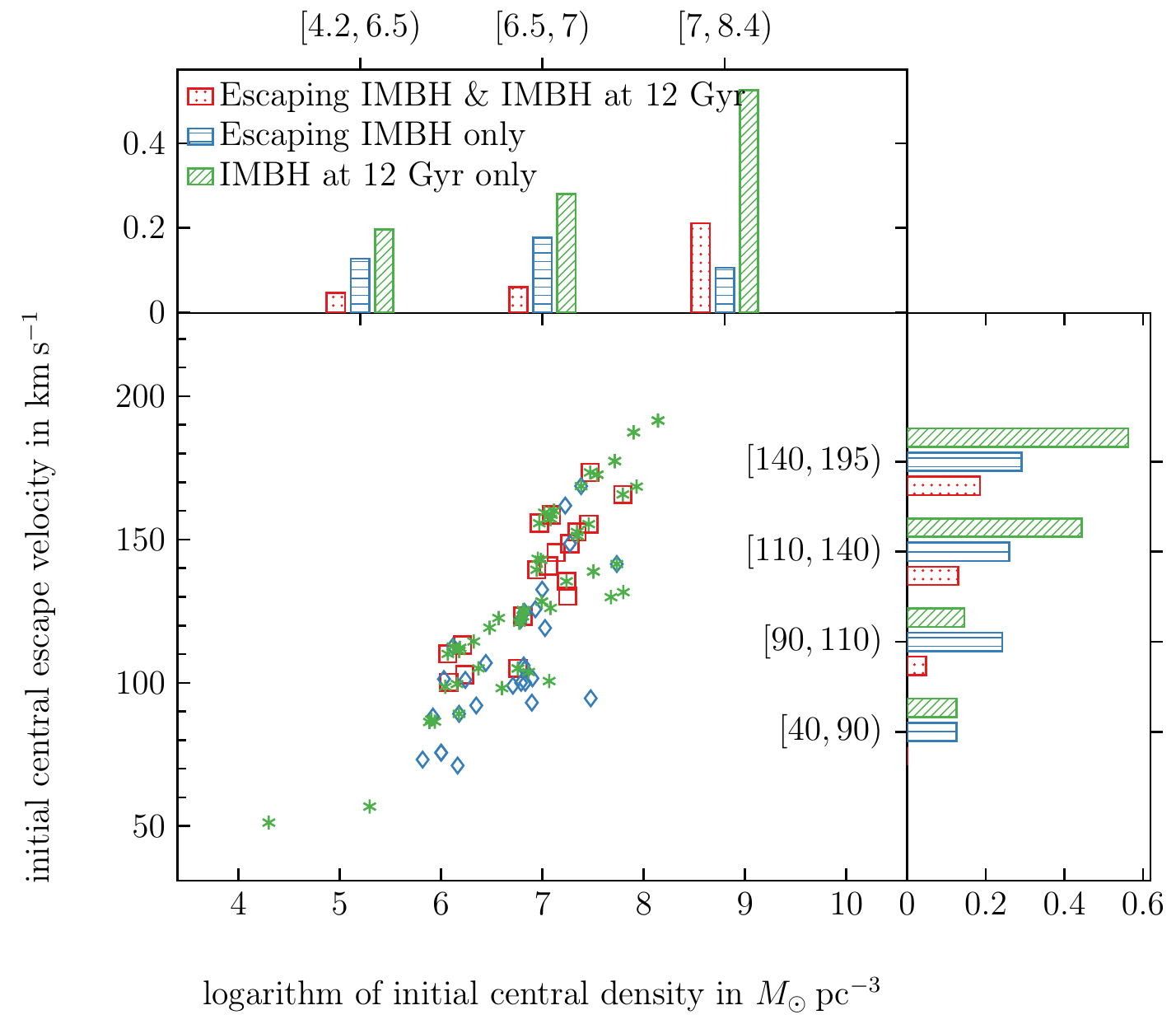}
    \caption{Initial central density \citep[as defined in][]{casertano} and initial central escape velocity plot. Side histograms show the share in the total number of models belonging to the same value range. Models where only IMBH escape have been observed are marked with red color (square marks and dotted histogram). Models with an IMBH present in the cluster at 12 Gyr and no escaping IMBHs during the evolution of the cluster are marked with green color (asterisk marks and oblique lines histogram). Those where both escape of an IMBH and its presence in the system at the Hubble time were observed are marked with blue color (diamond marks and horizontal lines histogram).}
    \label{fig:fig1}
\end{figure}

Both initial central density and initial central escape velocity, as expected from theoretical considerations, seem to influence the process of IMBH formation as well as those leading to its ejection from the cluster (see Fig.\,\ref{fig:fig1}). The higher the values of those parameters are, the higher the probability for an IMBH to form and stay in the cluster until 12 Gyr. For IMBHs that are escaping there is a peak for models with the values of logarithm of the initial central density ($\msun\:\textrm{pc}^{-3}$) between 6.5 and 7. This is caused by the fact that for larger densities the pace of dynamical interactions is higher (due to smaller relaxation time) and thus IMBHs acquire their mass faster and are becoming less susceptible to be removed.

\begin{table*}
\caption{Initial conditions for the MOCCA-SURVEY Database II models. A two segmented initial mass function (IMF) as given by \citet{2001MNRAS.322..231K} was used for all models, $f_\textrm{b}$ - binary fraction. If the binary fraction is equal to 0.95 then binary parameters are chosen
according to \citet{1995MNRAS.277.1507K}, otherwise eccentricity distribution is thermal, mass ratio distribution is uniform and semi-major distribution is uniform in logarithm, between $2(R_1 + R_2)$ and 100 AU. Most of the models (147 out of 212) have the $f_\textrm{b}=0.95$, the rest are models with $f_\textrm{b}=0.1$ (46) or $f_\textrm{b}=0.0$ (12). $N$ - initial number of stars and binaries, $R_{\textrm{t}}$ - tidal radius, $R_{\textrm{h}}$ - half-mass radius, $W_0$ - King model parameter, $Z$ - cluster metallicity. The upper mass limit for second generation stars was set to either $50$ or $150 \msun$, the lower mass limit is $0.08 \msun$. For each initial number of objects different combinations of parameters are used to generate the initial model.}
\label{tab:tab1}
\begin{tabular}{cccccccc}
    \hline
     $N$ & $R_{t}\:(\textrm{pc})$& $R_{\textrm{h}}\:(\textrm{pc})$ & $W_{0}$ & $Z$ & $f_{\textrm{b}}$ & Natal Kicks & \# of models  \\ \hline
     $6 \cdot 10^5$ & 60, 120 & 0.6, 1.2, 2.0, 4.0, 6.0 & 6, 8 & 0.001 & 0.1, 0.95 & Fallback & 66 \\
     $8 \cdot 10^5$ & 60, 120 & 0.6, 1.2 & 6, 8 & 0.001 & 0, 0.1, 0.95 & Fallback & 71 \\
     $10^6$ & 60, 120 & 0.6, 1.2 & 6, 8 & 0.001 & 0, 0.1, 0.95 & Fallback & 75 \\ \hline
\end{tabular}

\end{table*}

\begin{table}
\setlength\tabcolsep{0pt} 
\caption{Studied initial parameters of the models which constituted the survey. $N$ is the initial number of objects in the simulation, $\rho_{\textrm{c0}}$ is the initial central density expressed in $\rm M_{\sun}\:\textrm{pc}^{-3}$, $r_{\textrm{t0}}/r_{\textrm{c0}}$ is the concentration parameter (initial tidal radius to core radius), $r_{\textrm{t0}}$ is the initial tidal radius in $\textrm{pc}$, $v_{\textrm{ce0}}$ is the initial central escape velocity in $\textrm{km}\:\textrm{s}^{-1}$. For every value/range of the parameter the total number of models belonging to it is given.}
\label{tab:tab2}
\begin{tabular*}{\columnwidth}{@{\extracolsep{\fill}} lcccccc@{}}
\toprule
& \multicolumn{3}{c}{$N$} & \multicolumn{3}{c@{}}{$\log_{10}(\rho_{\textrm{c0}})$} \\ 
\cmidrule(lr){2-4} \cmidrule(l){5-7}
& \mc{$6\cdot10^5$} & \mc{$8\cdot10^5$} & \mc{$10^6$} 
& \mc{$[4.2,6.5)$} & \mc{$[6.5,7.0)$} & \mc{$[7.0,8.4)$}   \\ \midrule
total \# of models & 66     & 71     & 75     & 87      & 68     & 57          \\ 
\end{tabular*}
\begin{tabular*}{\columnwidth}{@{\extracolsep{\fill}} lccccc@{}}
\toprule
& \multicolumn{3}{c}{log$_{10}(r_{\textrm{t0}}/r_{\textrm{c0}})$} & \multicolumn{2}{c@{}}{$r_{\textrm{t0}}$} \\ 
\cmidrule(lr){2-4} \cmidrule(l){5-6}
& \mc{$[1.75,2.5)$} & \mc{$[2.5,2.75)$} & \mc{$[2.75,3.35)$} 
& \mc{$60$} & \mc{$120$} \\ \midrule
total \# of models & 68     & 81     & 63     & 140      & 72   \\    
\end{tabular*}
\begin{tabular*}{\columnwidth}{@{\extracolsep{\fill}} lcccc@{}}
\toprule
& \multicolumn{4}{c@{}}{$v_{\textrm{ce0}}$} \\
\cmidrule(lr){2-5}
& \mc{$[40,90)$} & \mc{$[90,110)$} & \mc{$[110,140)$}   & \mc{$[140,195)$} \\ \midrule
total \# of models & 48      & 62     & 54  & 48   \\    \bottomrule
\end{tabular*}
\end{table}

The concentration parameter comes from the \citet{1966AJ.....71...64K} model and it reflects the same dependency as the initial central escape velocity and initial density in the centre, which means that for an IMBH the higher the value of concentration parameter the more likely it will form. Our models are initially very dense (see Table \ref{tab:tab1} and \ref{tab:tab2}) and the process of the core collapse due to the mass segregation starts within the first few or few hundred Myr. After that the energy that is generated in dynamical interactions with the massive binaries starts to control the evolution of the system and the phase of evolution after the core collapse.

Additionally, we have checked whether IMBH formation or its ejection is dependent on the initial tidal radius and initial half-mass radius. We found out that both parameters should be checked on simultaneously and that IMBH ejection is not dependant on those parameters, but the formation is more likely to occur in models with higher tidal radius and lower half-mass radius. Such models can be treated as strongly tidally-underfilling and are well approximated by isolated system for the extended period of time, which means that they won't lose substantial part of their mass over given period of time.

Lastly, we found that our models match the theoretical predictions regarding the correlation between the central velocity dispersion and the possibility of IMBH formation due to mergers during few-body interactions \citep[see][]{2012ApJ...755...81M}. Within $66$ models in which an IMBH was present in the system at Hubble time and no previous IMBH have been observed only $5$ of them constituted cases where the velocity dispersion values were less than $40\:\textrm{km}\:\textrm{s}^{-1}$ (minimal value was $27 \  \textrm{km}\: \textrm{s}^{-1}$). The rest of them belonged to the range between $40$ and $87\:\textrm{km}\:\textrm{s}^{-1}$ which is in agreement with theoretical predictions for the velocity dispersion values in the centre of the cluster that is required to form a massive BH. The $40\:\textrm{km}\:\textrm{s}^{-1}$ velocity dispersion threshold is a value below which heating
from binaries prevents full core collapse. It is a result of a theoretical study which bases on assumptions and approximations regarding physical processes happening inside the cluster and thus cannot be treated as exact. 

\section{Escapers parameters} \label{sec:sec4}

There were in total $66$ escape events which included an IMBH and occurred in 49 models. The maximum of 3 IMBHs escaped from the single cluster. Most of all the ejection events have happened to the IMBH in a binary with another BH (60 out of 66). Two escapes concerned case in which both binary escapers were IMBHs. 

\begin{figure}
    \centering
    \includegraphics[width=\columnwidth]{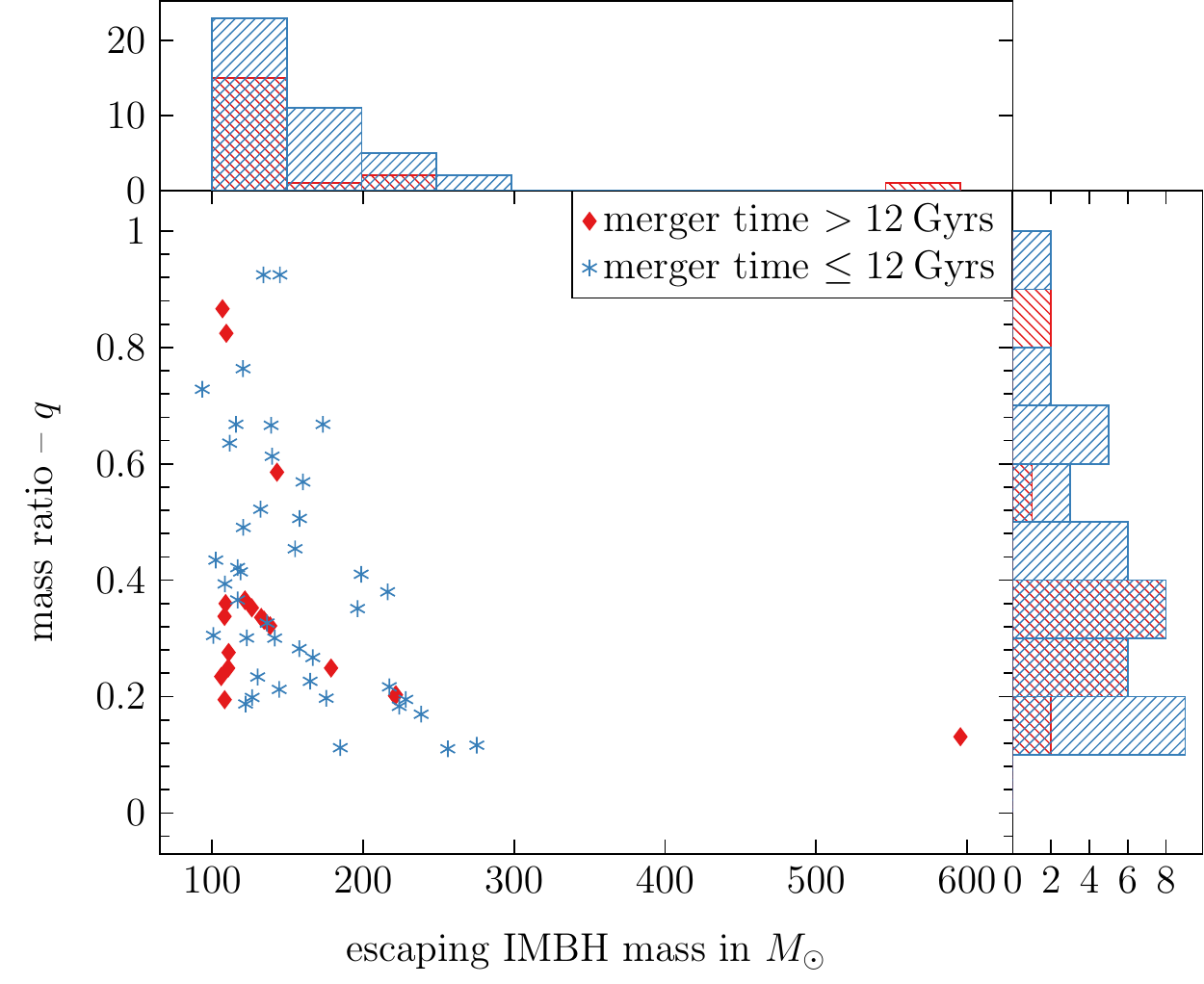}
    \caption{Mass ratio as a function of escaping IMBH mass. The histograms (to the right and above the scatter plot), show the frequency distribution of the mass ratio and escaping IMBH mass. 
    There were 6 single esacapers present in our models which are not plotted in this graph. Their masses were between $102$ and $115 \msun$. The red diamond mark represents binaries which will merge after 12 Gyr since the beginning of simulation, the blue asterisk mark is for the binaries that are going to merge after 12 Gyr.}
    \label{fig:fig2}
\end{figure}

In Fig.\,\ref{fig:fig2} we present the mass ratio of the escaping binaries as a function of the escaping IMBH mass. Single IMBH escapes and are not present in the plot. Masses of single escapers do no exceed $150 \msun$ whereas those that correspond to IMBH escapers in a binary can have masses that are twice as large. The mass ratio distribution of IMBH binary escapers with merger time less than 12 Gyr has two peaks: at around $0.1$ and at values between $0.3-0.4$ which corresponds to the companion mass of about $40 \ \msun$ and is similar to the mass of the most massive BH that can be formed via stellar evolution in the MOCCA-SURVEY Database II. Mergers occurring after 12 Gyr will most likely involve binary systems that have mass ratios between $0.2-0.4$. For most cases, masses of the escaping IMBHs do not exceed $300\:\msun$, however, a BH of nearly $600 \msun$ was ejected from the simulation and we present the exact history of this object in section \ref{sec:sec7}.

\begin{figure}
    \centering
    \includegraphics[width=\columnwidth]{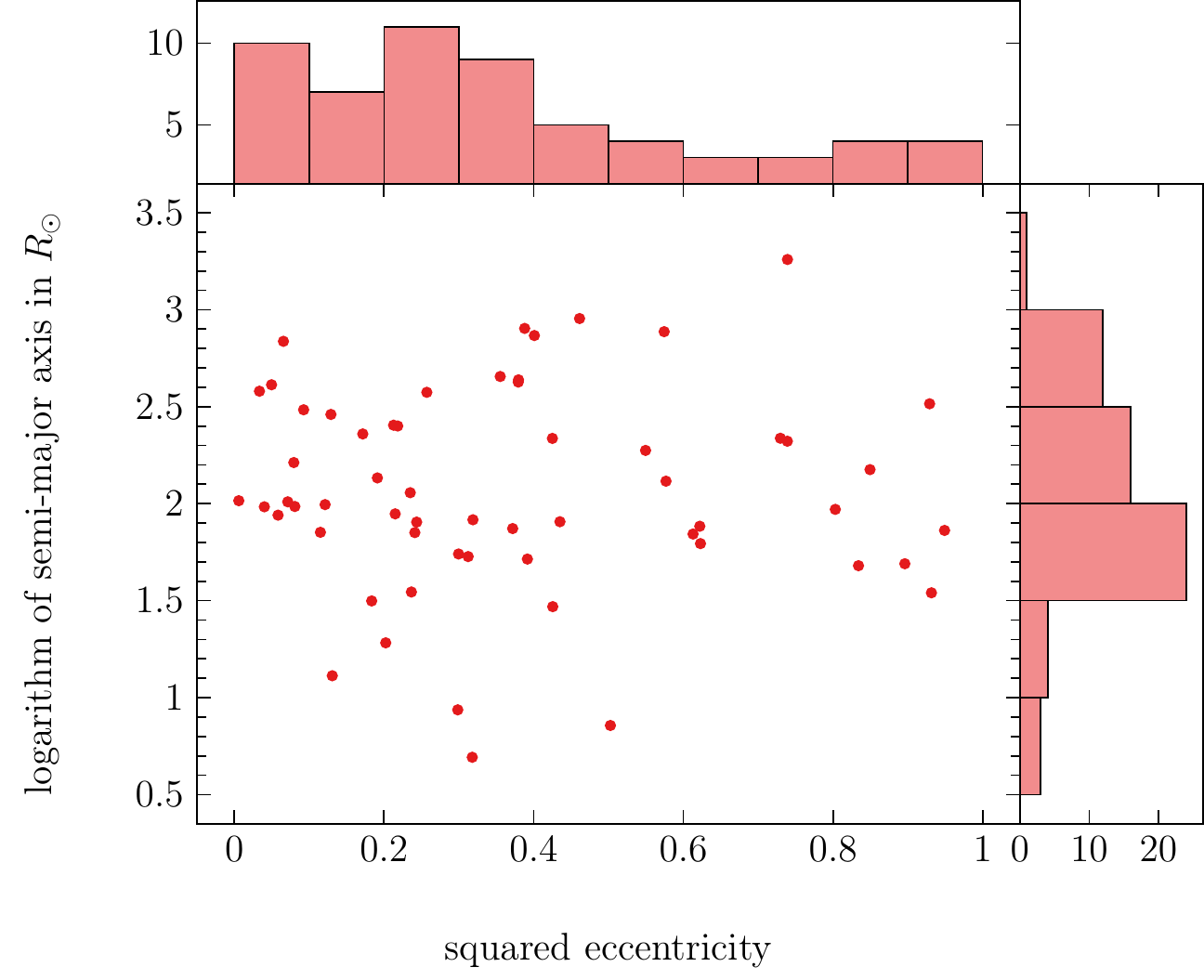}
    \caption{Semi-major axis logarithm and squared eccentricity plot of the IMBHs that are escaping in a binary. There are in total 60 of such cases. On the sides are plotted histograms with the number of occurrence as the $y$-value. Mean value of eccentricity equals $0.508$ and its dispersion is $0.277$.}
    \label{fig:fig3}
\end{figure}

Fig.\,\ref{fig:fig3} shows the semi-major axis (taken in logarithm) and squared eccentricity of the IMBHs escaping in a binary. The mean value of eccentricity is equal to $0.508$ with the dispersion of the mean value of $0.277$. After sufficient number of interactions, that the binary in a star cluster experiences, the eccentricity distribution should be thermal \citep[see][and references therein]{1975MNRAS.173..729H, Geller_2019}. Our data seems to be consistent with those theoretical predictions (mean value of eccentricity for thermal distribution is $2/3$) assuming the sigma test. This means that binaries that contain an IMBH have had a large number of interactions in the cluster before they managed to escape from it and are not primordial. It is also worth to note that there are no binaries with the low semi-major axis values and high eccentricities.
It comes from the fact that for such objects the time scale between mergers due to gravitational wave emission is shorter than the mean time between subsequent dynamical interactions. Most of the studied objects have the semi-major axis between $30$ to $100\:R_{\sun}$ and their distribution is not dependant on eccentricity for $\log{a}>1.5$.

\begin{figure}
    \centering
    \includegraphics[width=\columnwidth]{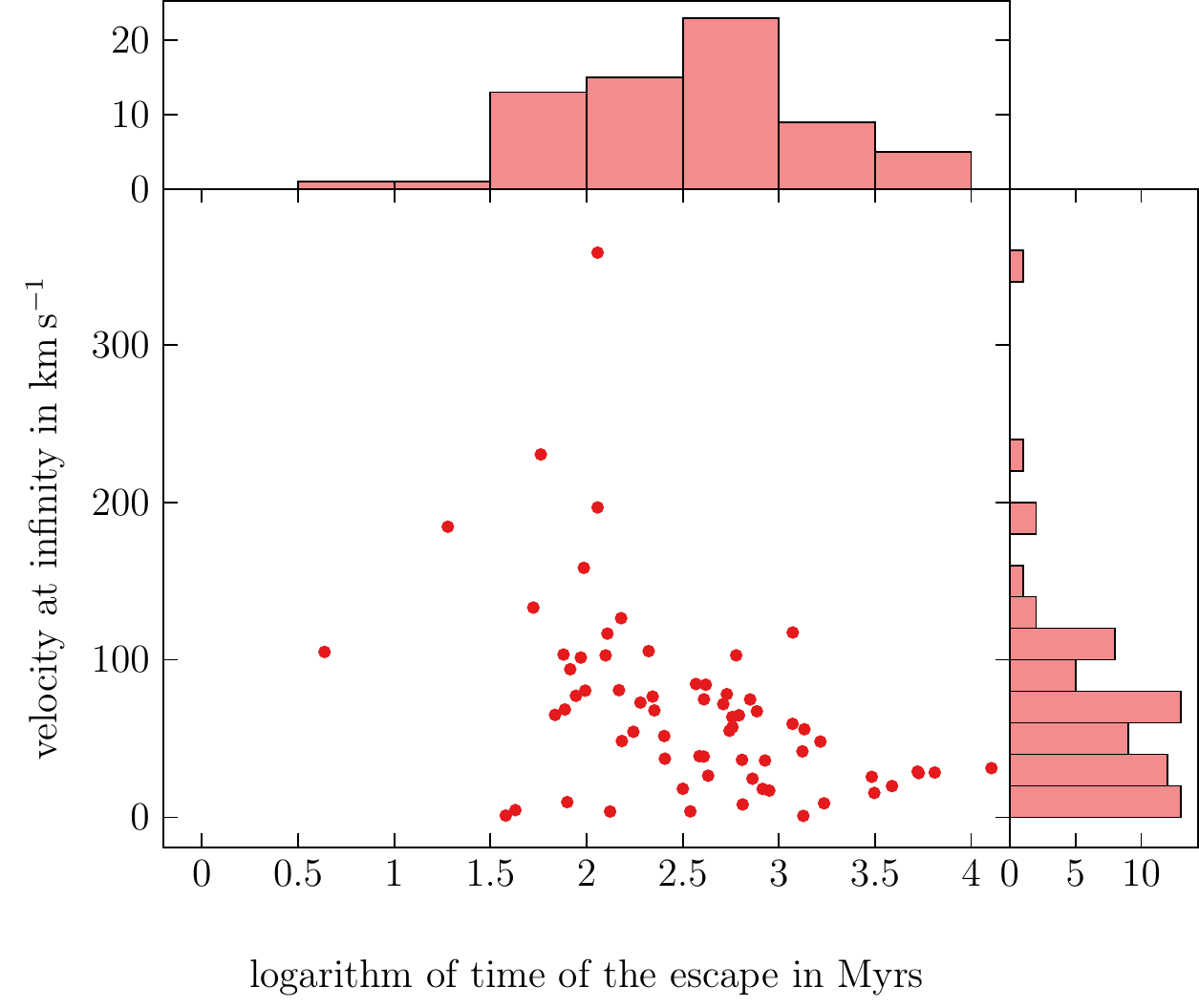}
    \caption{The velocity at infinity of the escaping IMBH (single or in a binary) as a function of the escape time in Myr taken in logarithm. On the sides are histograms that show the number of occurrence on the $y$-axis that correspond to different parameters.}
    \label{fig:fig4}
\end{figure}

Lastly, we present the velocity at infinity of the escaped object as a function of the time of the escape (taken in logarithm) in Fig.\, \ref{fig:fig4}. The velocity at infinity tends to get smaller as the time of the escape progresses. For a more dynamically evolved system there are objects with smaller average masses and thus the dynamical interactions become weaker. Additionally the half-mass radius is bigger as well as the central density is smaller, which lowers down the escape velocity. The values of the velocity are not high and in most cases they don't exceed $100\:\textrm{km}\:\textrm{s}^{-1}$. Taking into account the proper motion of the real  GCs \citep[e.g.][]{Dinescu_1999, 2021arXiv210808507B} we can conclude that substantial fraction of IMBHs that have escaped from the cluster is likely to be found in the vicinity of  GC tidal tails. 

\section{mass build-up events} \label{sec:sec5}

In order to determine how IMBHs that are escaping grew in mass we studied all mass build-up events that such BHs experienced from the time the BH seed have been formed up to its ejection as IMBH. We have put them into different categories to identify processes that lead to their mass increase. The result of the study is presented in Table \ref{tab:tab3}.

In Table \ref{tab:tab4} we have presented the same results for the IMBH that remained in the system and evolved until the Hubble time, in order to establish whether different processes are responsible for their mass build-up. 

\begin{table}
\setlength\tabcolsep{0pt} 
\caption{Table showing mass build-up events contribution to the total mass of escaping IMBHs. Dynamical events include: 3b,4b BH:BH -- three- or four-body encounter in which merger of two BHs occur, 3b,4b other -- three- or four-body encounter in which merger of a BH with object other than a BH (e.g. with a MS star, WD, NS) occurs and direct collisions (only with objects that are not BH). There were no collisions with other BHs. Evolutionary events consist of: BH:BH merger -- BBH merger, mass transfer in a binary and BH:other merger -- binary merger with object that is not a BH. Both total mass and mean mass are expressed in $M_{\sun}$.}
\begin{tabular*}{\columnwidth}{@{\extracolsep{\fill}} rlccc@{}}
\toprule
& & total mass & \# of events & mean mass \\ \midrule
\ldelim\{{3}{14.5mm}[\parbox{12.5mm}{dynamical}] & 3b,4b BH:BH & 484.7 & 11 & 44.07 \\ 
 & 3b,4b other & 115.9 & 48 & 2.41 \\
& other collision & 18.27 & 6 & 3.04 \\ \midrule
\ldelim\{{3}{14.5mm}[\parbox{12.5mm}{evolution}] & BH:BH merger & 4801 & 114 & 42.11 \\ 
& mass transfer & 38.55 & 10 & 3.85 \\
& BH:other merger & 245.2 & 12 & 20.44 \\ \bottomrule

\end{tabular*}
\label{tab:tab3}
\end{table}

\begin{table}
\setlength\tabcolsep{0pt} 
\caption{Table showing mass buildup events contribution to the total mass of IMBHs that are staying inside the cluster at 12 Gyr. Dynamical events include: 3b,4b BH:BH -- three- or four-body encounter in which merger of two BHs occur, 3b,4b other -- three- or four-body encounter in which merger of a BH with object other than a BH (e.g. with a MS star, WD, NS) occurs and direct collisions (both with BHs and objects other than a BH). Evolutionary events consist of: BH:BH merger -- BBH merger, mass transfer in a binary and BH:other merger -- binary merger with object that is not a BH. Both total mass and mean mass are expressed in $M_{\sun}$.}
\begin{tabular*}{\columnwidth}{@{\extracolsep{\fill}} rlccc@{}}
\toprule
& & total mass & \# of events & mean mass \\ \midrule
\ldelim\{{4}{14.5mm}[\parbox{12.5mm}{dynamical}] & 3b,4b BH:BH & 1435 & 39 & 36.8 \\ 
 & 3b,4b other & 19881 & 15626 & 1.27 \\
& BH collision & 129751 & 19162 & 6.77 \\
& other collision & 19611 & 28798 & 0.68 \\ \midrule
\ldelim\{{3}{14.5mm}[\parbox{12.5mm}{evolution}] & BH:BH merger & 21049 & 617 & 34.1 \\ 
& mass transfer & 41 & 906 & 0.05 \\
& BH:other merger & 226 & 308 & 0.73 \\ \bottomrule

\end{tabular*}
\label{tab:tab4}
\end{table}

We have found that IMBHs that are escaping build-up their mass most effectively through reapeted BBH mergers due to gravitational wave emission. In contrast those that are present in the system at 12 Gyr acquire their mass substantially through collisions (mostly with BHs). This is with agreement with the results from the MOCCA-SURVEY Database I \citep[see][]{2015MNRAS.454.3150G} that for a massive BH (over $\sim1000 \msun$) the physical collisions with other objects dominate the mass build-up process of an IMBH.

Firstly, IMBH seed will acquire its mass most effectively by the repeated BBH mergers. It happens in the very begining of the simulation when there is sufficient amount of BHs that constitute subsystem. While comparing the mean masses of the dynamical mergers between two BHs in Table \ref{tab:tab3} and \ref{tab:tab4}, it is clear that IMBH escapes happen at the very beginning of the simulation when evolutional mergers take place with most massive BHs present in the system. Mean increase of mass through BH mergers is also in agreement with the theoretical predictions, that firstly more massive BHs will be depleted from the BH subsystem. In this paper, however, we didn't study the influence of the gravitational recoil on the process. Due to anisotropy of masses and spins such mergers will gain velocity that can be greater than the escape velocity and in result can be ejected from the cluster. The code itself didn't account for the recoil so the situation in real cluster may be different. We studied the effect in the Section\,\ref{sec:sec8.5}.

If the BH is not ejected through dynamical interactions it will continue to acquire mass through collisions with other objects. It is then harder or even impossible to remove it from system both due to dynamical interactions or gravitational recoil kick as it will quickly deplete the population of stellar-mass BHs in the cluster \citep[see][]{2014MNRAS.444...29L,2015MNRAS.454.3150G}.

\section{Cause of the escape} \label{sec:sec6}

The purpose of the paper is to show the possibility of dynamical ejections of IMBHs from globular clusters as our models did not account for the gravitational wave recoil kick caused by the asymmetry of masses and spins of the merging BHs.
Studying IMBHs that are escaping from the cluster, we investigated the masses of interacting objects in the last encounter. In most cases the last interaction will be the one responsible for the ejection of our studied objects, which is connected with the fact that after a strong interaction binding energy will raise to the value which is sufficient for the object to escape from the cluster (it has to be greater than the tidal energy $-1.5GM/R_{\textrm{t}}$, where $M$ is the current cluster mass). In most cases the binding energy inside the cluster is positive, which means that an object is not bound to it.

\begin{figure}
    \centering
    \includegraphics[width=\columnwidth]{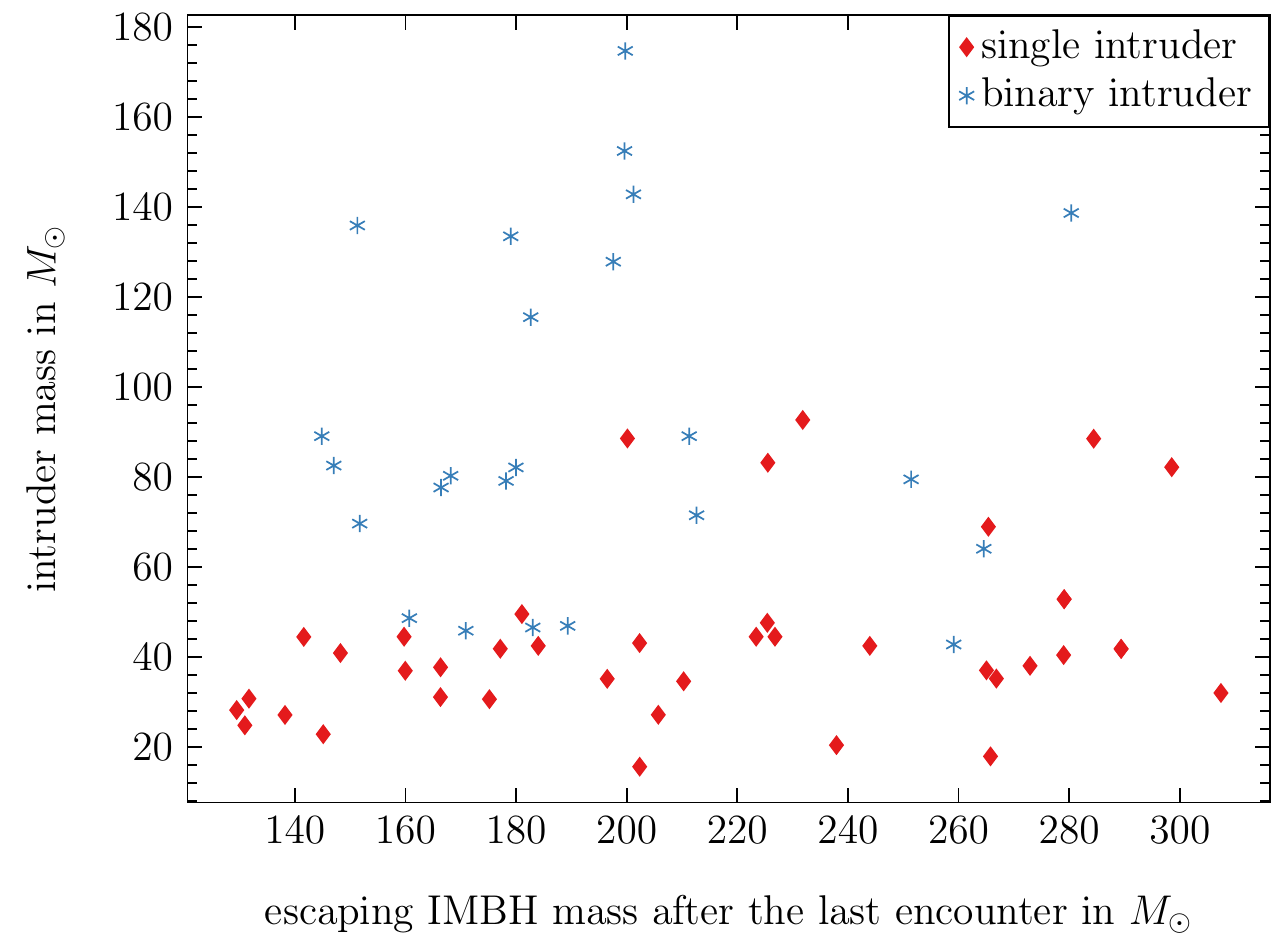}
    \caption{Escaping IMBH mass and intruder mass plot. Both masses are expressed in $M_{\sun}$. Escaping object is defined as single IMBH or a binary that involves an IMBH. Intruder is a single object (red color and diamond mark) or a binary (blue color or square mark) that interacts with the escaping object in the last encounter.}
    \label{fig:fig5}
\end{figure}

Subsequently, while studying the processes that can lead to an IMBH escape the key information is what are the masses of objects involved in the last encounter. In Fig.\,\ref{fig:fig5} we present both masses of an intruder and escaping object. In this case we define escaping object as a single IMBH or a binary involving an IMBH. Intruder is a single object or a binary that interacts with the escaping object in the last encounter. Two peaks of intruder mass at around $40$ and $80$ solar masses is caused by the fact that the most massive BH that can be formed in single star evolution is around $40 \msun$ in MOCCA-SURVEY Database II. This means that the intruder, in most cases, is a massive BH that did not further grow its mass, a binary consisting of two such BHs or their merger product.

It is worth noticing that the intruder mass does not have to be high for an IMBH to escape from the cluster. In fact, in most cases the mass of the intruder is 5-10 times smaller than the escaping object mass. This suggests that the binary increases its energy gradually and the final interaction is strong not due to the fact that the intruders mass is high but due to its really close approach to the very hard IMBH binary. 

\begin{figure}
    \centering
    \includegraphics[width=\columnwidth]{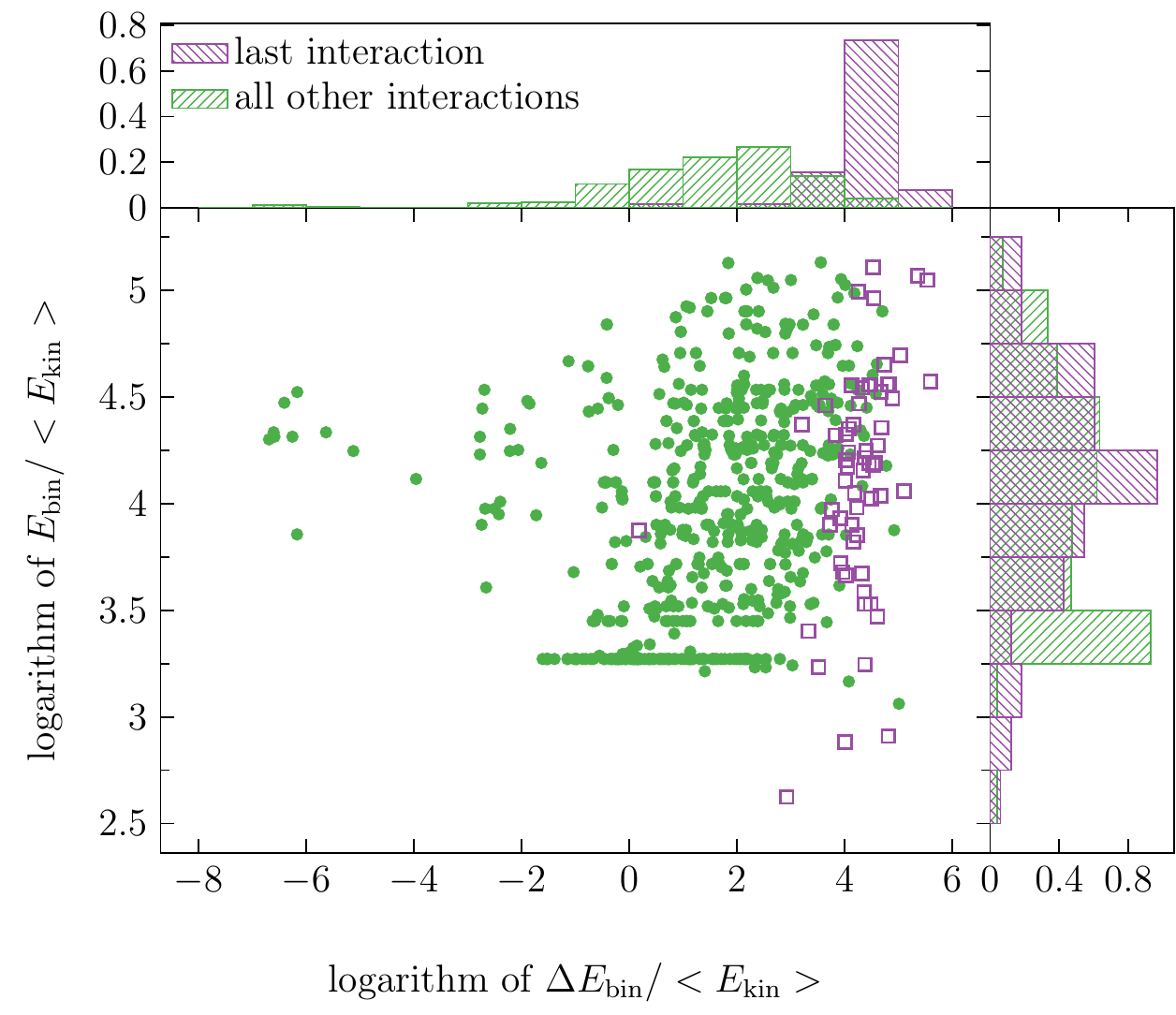}
    \caption{Binding energy of the escaping binary as the function of the binding energy change (both scaled to the mean kinetic energy) in the interaction. Plot was made for objects that were escaping in a binary and included only history of such objects from the time they were formed up to the last interaction. We have distinguished the last interaction with purple color (marked with dot), all previous are marked with green color (or square mark).}
    \label{fig:fig6}
\end{figure}

To support this claim, we have plotted the binding energy as the function of the binding energy change to mean kinetic energy of objects in the core for all 60 IMBHs that were escaping in a binary. Plots were obtained while studying the history of the binary that have escaped from the time it was formed up to the last interaction it took part in. We have distinguished the last interaction that they have experienced in the cluster and those that have happened before to compare them. Results are presented in Fig.\,\ref{fig:fig6} and clearly show that the energy changes in last interaction were comparable to the binding energy of the IMBH binary. Studied binaries are also very hard (their binding energy is over thousand times bigger than the mean kinetic energy in the core), which means that they must have encountered hundreds different interactions before being ejected from the cluster. This seems to support the picture that was resulting from the study of the masses of intruders. One interaction seems to stand out of all others and has the binding energy change at the same level as the mean kinetic energy in the core. This peculiar example is presented in section \ref{sub:sub71}. We finally studied the distance of closest approach for every last interaction that was a fly-by (28 cases) or exchange (15 cases) in a three-body encounter. We have found that in most cases, this distance is about hundredth part of a solar radius (calculated from the definition of the impact parameter: $b^2=r_p^2(1+2G(m_1+m_2+m_3)/(u^2r_p))$, where $r_p$ is the pericentre with respect to the centre of mass of the binary, $m_i$ are the masses of interacting objects and $u$ is the intruders relative velocity with respect to the centre of mass of the binary) and the intruder is always a BH. It means that these are very strong interactions that can significantly affect the orbit of the IMBH companion.

\section{IMBH escape examples} \label{sec:sec7}

\subsection{600 $M_{\sun}$ IMBH}\label{sub:sub71}

\begin{figure}
    \centering
    \includegraphics[width=\columnwidth]{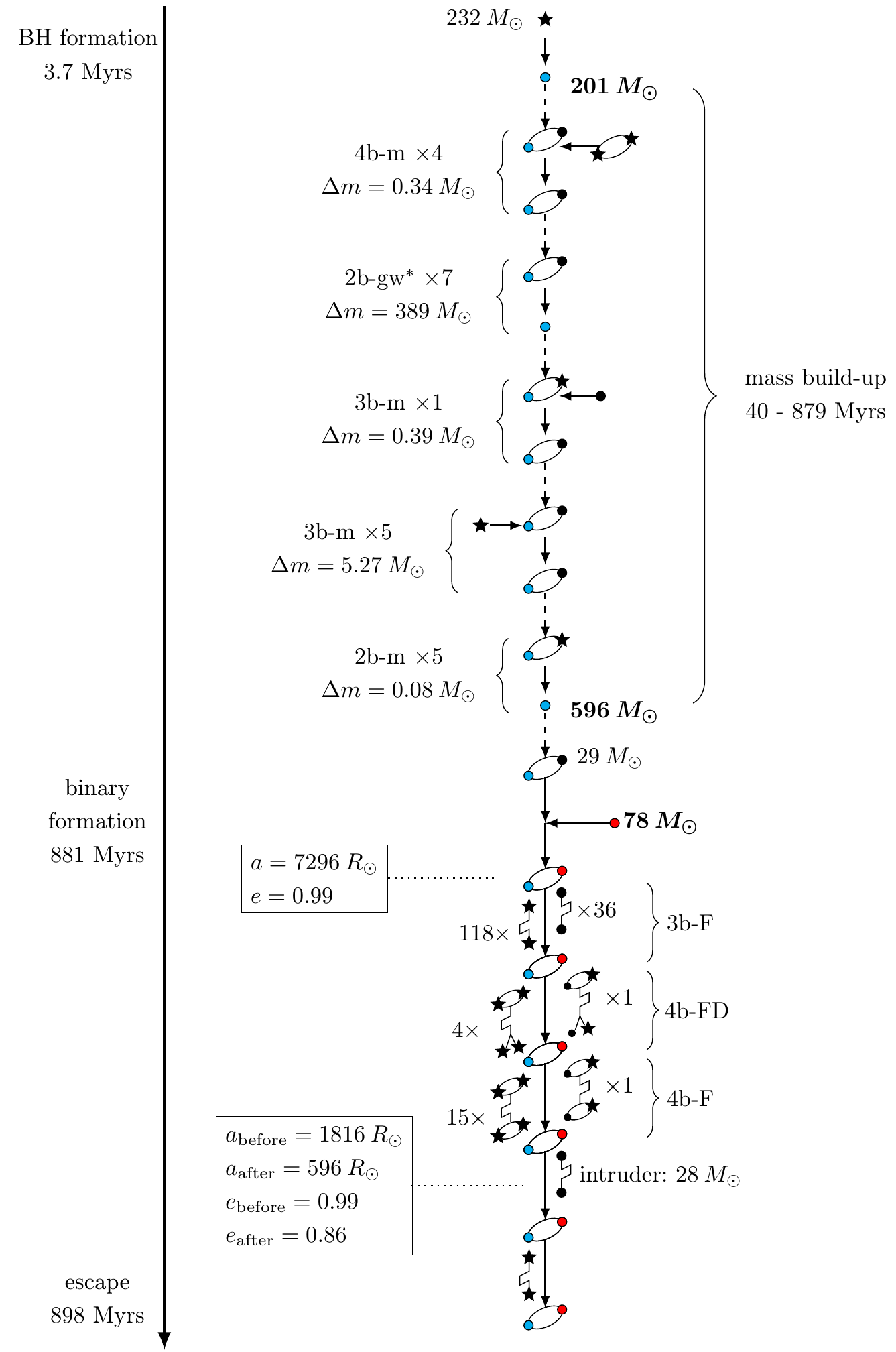}
    \caption{Diagram representing history of two BHs (blue color -- IMBH and red -- regular BH) escaping in a binary. BHs are marked with a dot whereas star represents any other object. A direct transition from one state of the system to another is marked with a solid line while dashed line is a non direct transition. Abbreviations used in the description of events: 3b-m -- coalescence due to three-body interaction, 2b-gw -- BBH merger due to the emission of gravitational waves, 4b-F -- four-body flyby, 4b-FD -- four-body flyby with a disruption of a binary, 3b-F -- three-body flyby.}
    \label{fig:fig7}
\end{figure}

Fig.\,\ref{fig:fig7} represents the history of nearly $600 \msun$ IMBH that escapes from the cluster. The initial parameters of the model are:
\begin{itemize}
    \item central density -- $6.94\cdot 10^6\:\rm M_{\sun}\:\textrm{pc}^{-3}$
    \item central escape velocity -- $122\:\textrm{km}\:\textrm{s}^{-1}$
    \item number of objects -- $6\cdot 10^5$
    \item tidal radius -- $60 \:\rm{pc}$
\end{itemize}
The initial parameters in this specific case correspond to the values which are on lower end of the values of both central escape velocity and central density.

Our given BH forms at $3.7$ Myr with $201 \msun$ as a result of a single star evolution of $232 \msun$ that has experienced multiple stellar mergers before. Such massive stars may form through stellar mergers or collisions as described in \cite{2021ApJ...908L..29G}. In the next phase it builds the mass through 19 different events that were presented graphically and are not in a chronological order. The biggest mass contribution in the final mass of a BH hold the BBH mergers that in 7 mergers acquired mass of nearly $400 \msun$. Our IMBH forms its final binary at $881$ Myr with semi-major axis of $7296\:R_{\sun}$ and $0.99$ eccentricity. Ensuing, it starts to harden due to interactions with other objects in the centre of GC and prior to the interaction responsible for its ejection, semi-major axis is lowered by the factor of $4$ which means that the binary binding energy increase was of the same level. It is worth mentioning that the interaction responsible for the ejection was not strong enough to remove the $600 \msun$ IMBH from the simulation immediately. The energy after the interaction was not sufficient for the object to escape, however, due to a large number of previous dynamical interactions the orbit of the object was highly eccentric. As a result, there was no need for strong interaction to eject it from the cluster because its binding energy was already greater than the tidal energy for the cluster.

\subsection{Two IMBHs binary} \label{sub:sub72}

\begin{figure}
    \centering
    \includegraphics[width=\columnwidth]{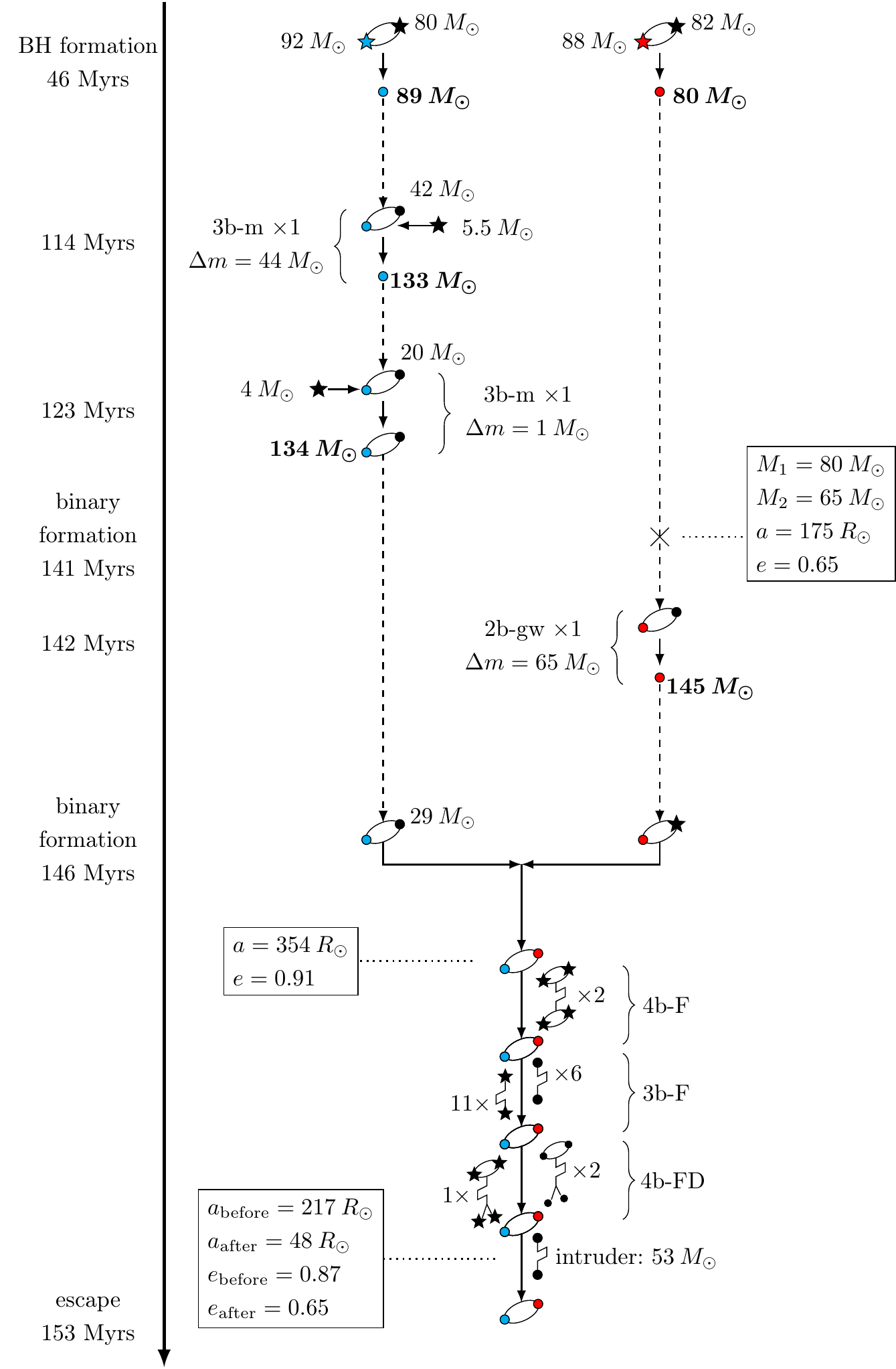}
    \caption{Diagram representing history of two IMBHs escaping in a binary. BHs are marked with a dot whereas star represents any other object. A direct transition from one state of the system to another is marked with a solid line while dashed line is a non direct transition. Abbreviations used in the description of events: 3b-m -- coalescence due to three-body interaction, 2b-gw -- BBH merger due to the emission of gravitational waves, 4b-F -- four-body flyby, 4b-FD -- four-body flyby with a disruption of a binary, 3b-F -- three-body flyby.}
    \label{fig:fig8}
\end{figure}

Another interesting example of the escaping IMBHs are the ones that are escaping together as a binary. In Fig.\,\ref{fig:fig8}, we present one out of two such cases that occured in our simulated models. 
The initial parameters of the model are:
\begin{itemize}
    \item central density -- $2.41\cdot 10^7\:\rm M_{\sun}\:\textrm{pc}^{-3}$
    \item central escape velocity -- $169\:\textrm{km}\:\textrm{s}^{-1}$
    \item number of objects -- $10^6$
    \item tidal radius -- $120 \:\rm{pc}$
\end{itemize}
Both of the BHs form roughly at the same time in the evolution of the binary star system. One of the BHs forms with $89 \msun$ and builds its mass in two events. First it merges with another BH and a star in a three-body interaction, after which it forms a binary again and due to three-body interaction a star merges directly into it. The second BH forms with $80 \msun$ and gains its mass solely due to BBH merger. They both form a binary together at $146$ Myr and acquire energy needed to escape by the interactions in the centre of GC with other objects. The one responsible for the ejection of a binary is a flyby of $53 \msun$ intruder.

\section{Gravitational kicks} \label{sec:sec8.5}

\begin{figure}
    \centering
    \includegraphics[width=\columnwidth]{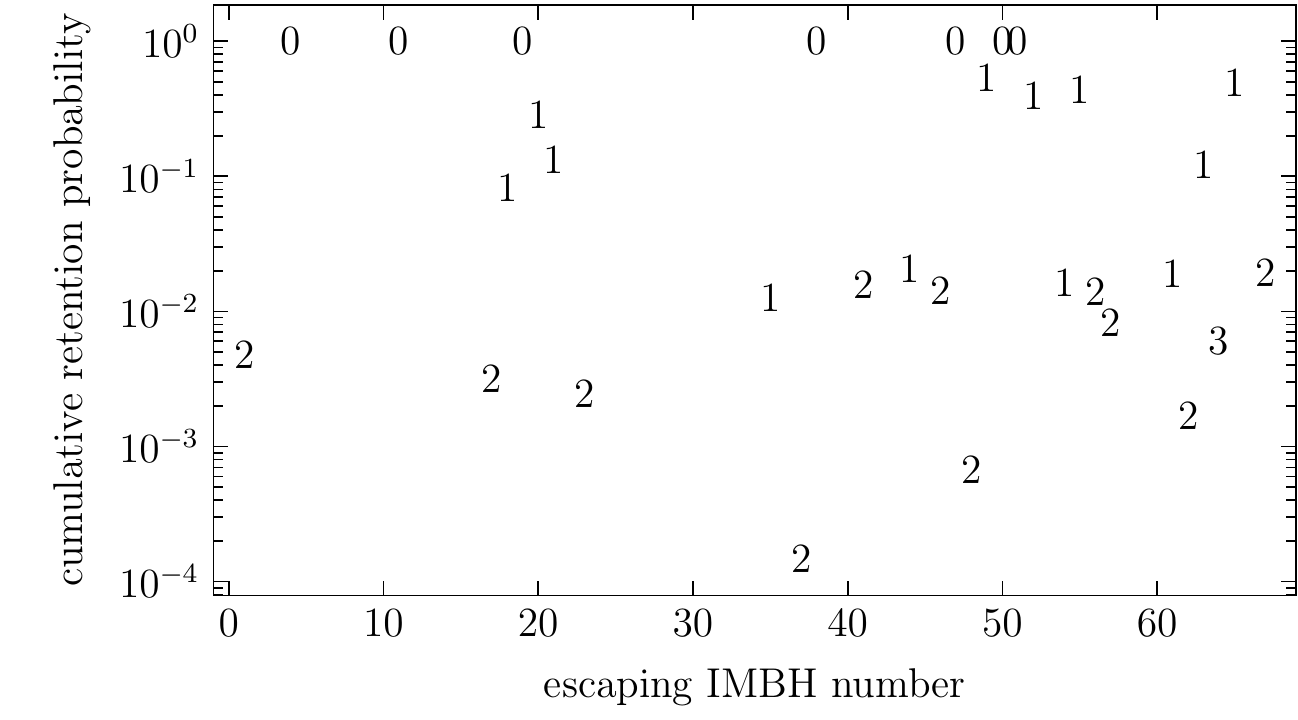}
    \caption{The overall (cumulative) retention probability for each of the IMBH escapers. The probabilities are marked with a number corresponding to the length of the merger chain that given IMBH was part of. The initial spin model is the constant spin model. In the plot are present only merger chains for which the overall retention probability was greater than 0. These include: 9 chains with length of 1, 15 chains with length of 2, 4 chains with length of 3, 6 chains with length of 4, 2 chains with length of 5, 1 chain with length of 7.}
    \label{fig:fig9}
\end{figure}

\begin{figure}
    \centering
    \includegraphics[width=\columnwidth]{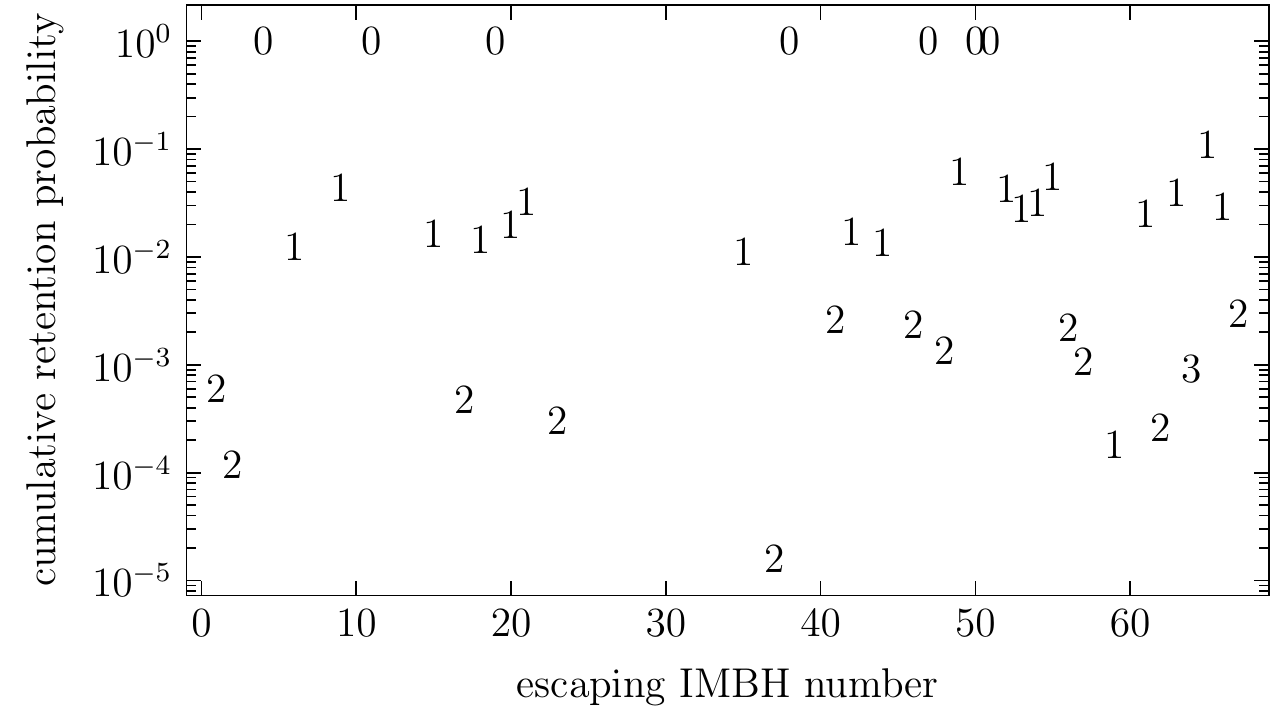}
    \caption{Similar to Fig.\,\ref{fig:fig9}, we show the overall (cumulative) retention probability for each of the IMBH escapers. The probabilities are marked with a number corresponding to the length of the merger chain that given IMBH was part of. The initial spin model is the random spin model. In the plot are present only merger chains for which the overall retention probability was greater than 0. These include: 2 chains with length of 1, 14 chains with length of 2, 4 chains with length of 3, 6 chains with length of 4, 2 chains with length of 5, 1 chain with length of 7. }
    \label{fig:fig10}
\end{figure}

In this section we will present our predictions regarding the influence of the gravitational kicks on our results. We tried to quantify it by calculating the retention probability for each BBH merger that involved our studied IMBHs. The gravitational kick prescriptions were taken from \citet{2018MNRAS.481.2168M}. These are the steps we used to calculate the retention probability for each merger:

\begin{enumerate}
    \item First the initial spin model had to be chosen between:
    \begin{enumerate}
        \item Constant spin model with the initial value of spin $0.1$. The assumed value is motivated by the gravitational wave observations \citep{gwtc-2, nitz} and is thought to represent the small value of the observed BBH mergers components spin.
        \item Random spin model -- Gaussian distribution of the spin centered around $0.5$ and a cutoff at $0.5/3$. The purpose of this model is to study the broader range of the initial spin values and it covers nearly the whole range of the spin magnitude.
    \end{enumerate}
    \item Next we drew both spin orientations $10^5$ times by drawing them uniformly on a sphere of a radius corresponding to the magnitude of the spin. For each spin configuration we calculated the recoil velocity of the merger product and compared it with the GCs central escape velocity at that time. Nearly all of our studied mergers were the case of the dynamically formed BBH systems (with only 1 exception of the primordial one) and is the reason for the assumed spin orientation treatment. The number of cases in which the recoil velocity was smaller than the central escape velocities divided by the total number of configurations is the retention probability for the BBH merger product.
    \item Previous point was repeated 10 times and the average as well as dispersion of the results had been calculated. From now on we will refer to the mean of the retention probabilities as being the retention probability of the merger product. 
    \item For each merger chain in the studied IMBHs history, we calculated the overall retention probability of the whole chain of mergers by multiplying probabilities for each node (merger event). These overall probabilities, marked with a number corresponding to the length of the merger chain, are shown in Fig.\,\ref{fig:fig9} (constant low spin model) and Fig.\,\ref{fig:fig10} (random spin model).
\end{enumerate}

For the constant (and low) initial spin model the overall cumulative retention probability is higher for IMBHs that have encountered 1 BBH merger. This comes from the fact that the first BBH merger involves BHs of a similar mass and their mass ratio is close to 1. The recoil velocity of the merger product for low spins (both spins having same magnitude) and high mass ratio is highly dependant on the spin orientations and its dispersion is large. If on the contrary the spins differ and not necessarily take low values the recoil velocity is very high compared to the central escape velocity and thus the retention probability will be low.

It is also worth noticing that not the first merger but the second one is the one that will most likely break the merger chain before the BH will gather enough mass to become IMBH. It comes from the fact that the first merger involves components of similar mass and thus their mass ratio is close to 1. After merging the outcome mass will be more less twice higher and it will experience higher recoil velocity due to higher mass asymmetry \citep[there is a peak in the mean recoil velocity for $q\approx 0.6$ -- see][]{2018MNRAS.481.2168M} during the next merger, with different BH.

Nevertheless, our post processing study shows that there are still some BHs that will form IMBHs, which will later on be ejected from the cluster solely due to dynamical interactions inside it. Depending on the assumed initial spin model there are about 7 -- 15 BHs (10 -- 22\% of all IMBH escapers) that will be removed from the cluster in the way presented in this paper.
\normalfont

\section{Discussion and conclusions} \label{sec:sec9}

In this work we explored the possibility of IMBHs being ejected from the star cluster solely due to dynamical interactions. Using a set of 212 dense and strongly tidally-underfilling star cluster simulated as part of the preliminary MOCCA-SURVEY Database II, we studied how different initial parameters affect the process. We found that IMBHs are most likely to be ejected in models with initial central densities between $\sim 3\cdot10^6 - 10^7  \msun \ \rm{pc}^{-3}$. For larger densities, the rate of dynamical interactions is higher and thus IMBHs acquire their mass faster and are less susceptible to be removed from the cluster. On the other hand, for lower densities, dynamical encounters are less frequent and such models are likely to never form an IMBH \citep{Hong2020MNRAS.498.4287H}. 

In total we observed 68 IMBH escapers coming from our models and most of them have a mass lower than $200 \msun$. Due to the fact that a significant number of our models had large initial central densities and additionally the code itself did not account for the gravitational recoil kicks of the merging BBHs, our results must be treated as the upper limit estimates of the process. Moreover, our models represent GCs which are rather dense and in many observed clusters the initial central densities may be lower, which will result in a lower probability of forming an IMBH and a smaller number of escapers.  We estimated that around 10 - 22 \% of the escapers present in the Survey will be subject to this dynamical ejection scenario, whereas the rest will be removed from the cluster due to gravitational kicks from the merger with another BH. Escaping IMBHs acquire their mass most effectively through BBH mergers and thus are likely to be removed before they reach $100\:M_{\sun}$ due to gravitational wave recoil kicks \citep{gerosa2019}. One of the concerns regarding gravitational kicks and their influence on the cluster evolution could be that the BH population is depleted and thus will not allow the IMBHs formation or that there will not be enough dynamical interactions with massive objects to eject them from GCs as we presented here. The number of BH:BH mergers is however small (order of tens) compared to the number of unique BHs in our models (order of thousands) and will not strongly affect the clusters evolution. Even when the merger product remains in the cluster its orbit can be changed into more elongated one. This does not necessary imply that it will no longer be subject to dynamical interactions in the core as its barycenter will be positioned there and every orbital period the BH will pass through the dense core where the probability of dynamical interactions is substantial due to its large mass.

The tidal disruptions were not considered in the code, however, the effect of the mass increase for dynamically escaping IMBHs can be neglected. Roche distance is (for mean mass escaping IMBH and MS star of 1 stellar mass): $d=R_{\sun}(M_{IMBH}/M_{\sun})^{1/3}\approx4.9R_{\sun}$. In the code the threshold for calculating the collision probability is $R_{\sun}$. This means that the mass increase effect would be even smaller than that and combined with the picture resulting from Table\;\ref{tab:tab3}, we conclude that this additional contribution to the total mass would still be insignificant compared to the mass gained through BH:BH mergers.

In the post processing of the data from our simulated stellar cluster models, we calculated the retention probability of merged BHs assuming different values of initial spin magnitudes and orientations. The spins were  drawn from a random uniform distribution and we took $10^5$ different orientations. For every set of spin orientations, kick velocities were calculated from prescriptions taken from \citet{2018MNRAS.481.2168M}.  We found that if we assume the initial spin magnitudes to be low (we assumed it to be 0.1, which is in fact consistent with observational evidence - \citet{2021PhRvX..11b1053A, 2021arXiv210509151N}), the retention probability becomes relatively high (around 0.4) for mass ratios close to 1, which corresponds to the first merger. This is caused by the fact that for low spin model, the kick velocity has its peak at around $q=0.4$ and is approximately constant for higher $q$ values. In addition to this, for higher $q$ the spin orientation starts to play key role for the kick velocity as the mean kick velocity dispersion grows. The problem of retaining the merger product is more complicated for repeated mergers. On one hand, each merger will gradually increase the spin magnitude and kick velocities will become really high. On the other, the build-up of IMBH mass will gradually lower down the value of $q$. It is also worth noting that the mass build-up of an IMBH is connected with mass accretion in collisions and with mergers with stars, which will result in an increase of the spin of a given IMBH. 

Binaries play key role in dynamical processes that lead to an IMBH ejection as they are able to acquire substantial amount of energy needed for those massive objects to escape from star cluster. The reason of escape is very close approach (hundredth parts of solar radii) of the intruder that can significantly affect the orbit of the IMBH companion.

Finally, we calculated the merger rate for IMBHs that have escaped in a binary (merger rate outside the cluster for IMBH binary) $\mathcal{R}=0.7\:\textrm{Gpc}^{-3}\:\textrm{yr}^{-1}$. This value is the upper limit for merger rate estimation as our models are very dense and the effect of the gravitational kick due to BBH merger has not been accounted for.

\section*{Acknowledgements}

MG, AH and AA were
partially supported by the Polish National Science Center (NCN) through the grant UMO-2016/23/B/ST9/02732. AH is also supported by the Polish National Science Center
grant Maestro 2018/30/A/ST9/00050. DGR and AA acknowledge support through the COST action CA16104. DGR was partially supported by the NCN grant UMO-2017/26/M/ST9/00978 and by POMOST/2012-6/11 from the Program of Foundation for Polish Science co-financed by the European Union within the European Regional Development Fund.
AA is supported by the Swedish Research Council through the grant 2017-04217.

KM would like to express his gratitude towards MG whose patient supervision has made publishing this work possible.  

\section*{Data Availability}

The data underlying this article will be shared on reasonable request to MG or AH.



\bibliographystyle{mnras}
\bibliography{bibliography} 








\bsp	
\label{lastpage}
\end{document}